# Conduction mechanism and switchable photovoltaic effect in (111) oriented BiFe$_{0.95}$Mn$_{0.05}$O$_3$ thin film

J. Belhadi[1], J. Ruvalcaba[2], S. Yousfi[1], M. El Marssi[1], T. Cordova[2],
S. Matzen[3], P. Lecoeur[3], and H. Bouyanfif[1]

[1]LPMC EA2081, Université de Picardie Jules Verne 33 Rue Saint Leu, 80000 Amiens, France
[2]División de Ciencias e Ingenierías, Universidad de Guanajuato campus León, México
[3]Centre de Nanosciences et de Nanotechnologies, Univ. Paris-Sud, CNRS, Université Paris-Saclay, Orsay, France

**Abstract**

Epitaxial 200nm BiFe$_{0.95}$Mn$_{0.05}$O$_3$ (BFO) film was grown by pulsed laser deposition on (111) oriented SrTiO$_3$ substrate buffered with a 50nm thick SrRuO$_3$ electrode. The BFO thin film shows a rhombohedral structure and a large remnant polarization of Pr = 104 μC/cm$^2$. By comparing I(V) characteristics with different conduction models we reveal the presence of both bulk limited Poole-Frenkel and Schottky interface mechanisms and each one dominates in a specific range of temperature. At room temperature and under 10mW laser illumination, the as grown BFO film presents short-circuit current density (J$_{sc}$) and open circuit voltage (V$_{oc}$) of 2.25mA/cm$^2$ and -0.55V respectively. This PV effect can be switched by applying positive voltage pulses higher than the coercive field. For low temperatures a large V$_{oc}$ value of about -4.5V (-225kV/cm) is observed which suggests a bulk non-centrosymmetric origin of the PV response.

## I. Introduction

During the last decade, multiferroic (MF) perovskite materials have been studied extensively due to the co-existence of ferroelasticity, ferroelectricity and magnetism in a single phase suitable for many advanced applications [1,2]. In addition, due to their relatively small bandgap (below 2.7eV) some MF materials have attracted attention for photovoltaic (PV) applications [3–6]. Actually BiFeO$_3$ (BFO) appears as the most extensively investigated MF material, since it exhibits room temperature ferroelectric (T$_C$ ~1103 K) and G-type antiferromagnetic (T$_N$ ~640 K) orders interesting for magnetoelectric coupling (magnetic order can be controlled by an electric field or vice versa) [7,8]. Bulk BFO has a R*3c* rhombohedral symmetry at room temperature (RT) ($a_{pc}$=3.965 Å and $α_{pc}$=89.3–89.5°) and a bandgap of about 2.7eV [7,8]. Its robust polarization (100μC/cm$^2$) and relatively small bandgap compared to the typical ferroelectric materials (BaTiO$_3$ and Pb(Zr,Ti)O$_3$) have attracted great attention for PV applications [5,9]. In 2010 Yang *et al.* reported a peculiar photovoltaic effect with a large open circuit voltage V$_{oc}$ of 16V in epitaxial BFO film using an in-plane geometry of measurements [9]. The obtained anomalous PV effect in BFO was explained by photovoltage generation at nano-scale ferroelectric domain walls (DWs) [9]. Three years later, Bhatnagar *et al.* observed the same anomalous PV effect with V$_{oc}$ above the band gap on a BFO film and ruled out the interpretation given by Yang *et al.*. Bhatnagar *et al.* explained instead the large V$_{oc}$ values by a non-centrosymetric bulk effect [5]. Such bulk photovoltaic effect is also observed in non-centrosymmetric perovskite materials such as LiNbO$_3$ and BaTiO$_3$ [10–12]. The origin of the PV effect in ferroelectric materials is strongly dependent on the microstructure, the geometry of measurements and the nature of the electrode/film interfaces. Numerous reports highlighting the contribution of defects show the complexity of the subject [13–15]. A clear understanding of the influence of ferroelectric polarization, defects, conduction mechanisms and film/electrode interface characteristics is obviously a prerequisite prior to any applications.

In this paper, we report on the investigation of the conduction mechanism and photovoltaic effect of a high quality epitaxial (111) oriented BFO thin film sandwiched between ITO top electrode and SRO bottom electrode. (111) orientation is chosen because of the bulk [111] BFO spontaneous polarization. The conduction mechanisms were studied by fitting the temperature dependence of the I(V) characteristics by different models including bulk limited and interface limited transport mechanisms. To gain insights on the PV mechanism the photo-induced effects were studied in a broad range of temperature in the as grown film and by applying voltage pulses with different sign to the ITO/BFO/SRO heterostructure.

## II. Experimental details

(200nm)BFO/(50nm)SRO heterostructure was grown on (111) oriented SrTiO$_3$ (STO) single crystal substrate by pulsed laser deposition (PLD) using a KrF laser (248nm). The fluence and pulse frequency of the laser were fixed respectively at 1.6J/cm$^2$ and 4Hz. BFO was grown under 5.10$^{-2}$ mbar of oxygen pressure (PO$_2$) at temperature of 725°C while SRO layer used as a bottom electrode was grown under 0.3mbar of PO$_2$ and 710°C. Bi$_{1.1}$Fe$_{0.95}$Mn$_{0.05}$O$_3$ target was used with 10% excess of Bi and 5% doping of Mn on the B site to respectively prevent Bi vacancies and lower the leakage currents [16,17]. The structure was characterized using a high-resolution 4 circles diffractometer (Bruker Discover D8). The electrical measurements were performed using Indium Tin Oxide (ITO) as top electrodes (0.1mm and 0.2mm diameters) deposited by PLD at room temperature under 1.6×10$^{-2}$ mbar PO$_2$ through a shadow mask. The P-V loops measurements were performed in the dynamic mode at 1kHz using a TF2000 analyzer while the I(V) characteristics were collected using a Keithley 2635 electrometer. The PV measurements were realized by illuminating the sample with an Argon-Krypton tuneable laser. The temperature was controlled using a Linkam stage that allows a temperature stability of 0.1K. All electrical measurements have been done by applying the voltage to the SRO bottom electrode.

## III. Results and discussions

Figure 1(a) displays room-temperature high resolution x-ray diffraction (XRD) pattern of BFO film with a thickness of about 200nm grown on (111) STO substrate buffered with a 50nm thick SRO conducting layer. Only the *(111)* and *(222)* diffraction peaks of BFO, SRO and STO are present showing the single orientation of the heterostructure with no impurity phase. The inset of Fig.1(a) shows the rocking curve (ω-scan) performed on (111) BFO, SRO and STO peaks. The full width at half maximum (FWHM) of BFO, SRO and STO are 0.28°, 0.18° and 0.05 respectively, indicating good crystallization of the BFO and SRO films with low mosaicity. Phi-scan measurements (Fig.1(b)) obtained for BFO film and the substrate on the (020) family of plane showed three peaks equally separated by 120° and at the same angle for both BFO and STO. This confirms the epitaxial growth of the BFO film and demonstrates an in-plane lattice alignment between the film and the substrate ([11-2]$_{BFO}$//[11-2]$_{STO}$ and [1-10]$_{BFO}$//[1-10]$_{STO}$).

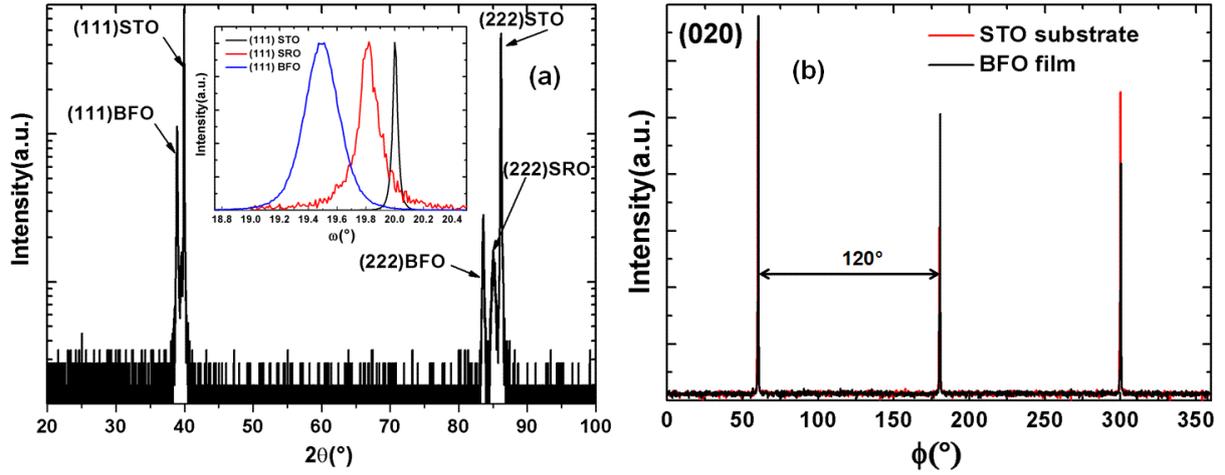

*Figure 1: (a) XRD pattern of (111) BFO/SRO/STO heterostructure (b) Phi-scan measurements on the (020) for BFO and SRO layers and the STO substrate. The insert in Fig.1(a) shows the ω-scan performed on first order of BFO, SRO and STO.*

From 2θ value of (222) peaks, the out-of-plane interplanar spacing of the BFO and SRO layers was found to be 2.31Å and 2.27Å, respectively. These values are higher than their corresponding BFO and SRO bulk parameters [18]. This increase is probably due to the in plane compressive strain applied by the substrate and/or to the presence of oxygen vacancies.

Figure 2 shows the P-V hysteresis loops and the current measured on the ITO/(111)BFO/SRO heterostructure at frequency of 1kHz. A saturated loop was obtained confirming the ferroelectric character of BFO film and the switching of the polarization. A large value of the remnant polarization is obtained ($P_r = 104$ μC/cm$^2$) which is little higher than that measured previously on (111) oriented BFO film [18].

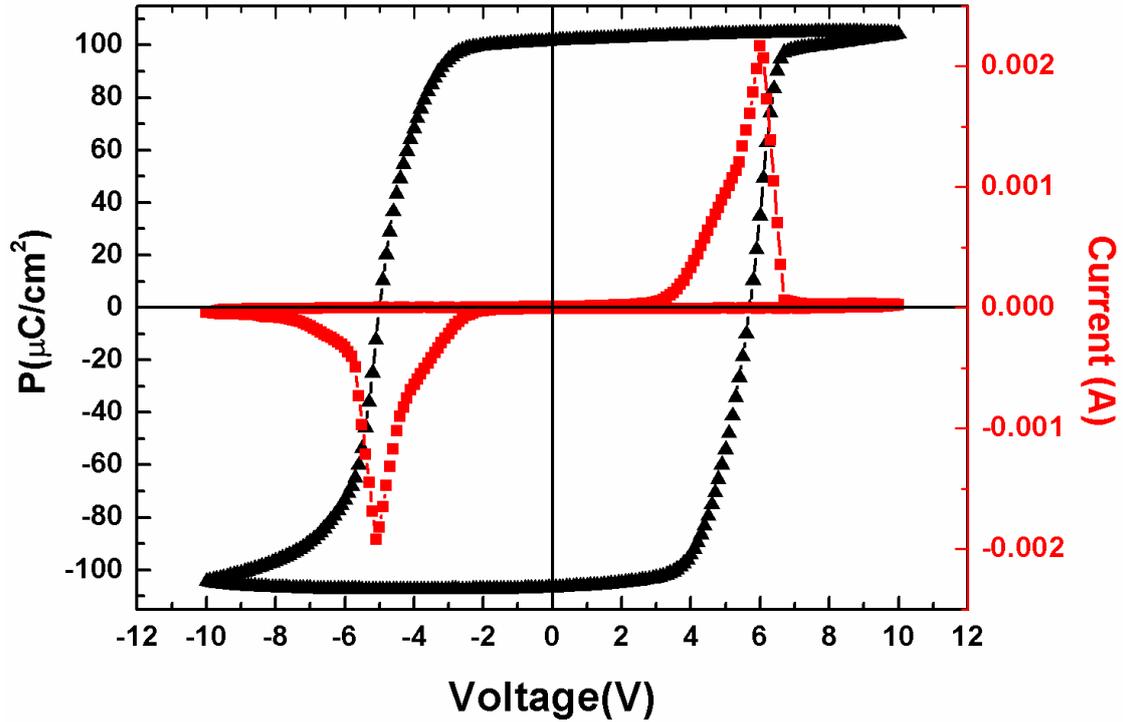

*Figure 2: P-V hysteresis loops and the current measured on the ITO/BFO/SRO capacitor.*

The shape of the hysteresis loop is rectangular indicating a good quality of BFO film and a robust ferroelectric state. The measured P-V loop presents an imprint (shift to the positive field) with a positive coercive voltage of 5.7V and a negative coercive voltage of -5.0V. This phenomena is frequently observed in BFO films and was related to the built-in electric field [14,17]. Such imprint can be explained by the use of different electrodes, strain gradient and/or a gradient of oxygen vacancies related to the growth of the BFO film [14].

Figure 3 shows the I(V) characteristics of ITO/BFO/SRO capacitor measured in dark at different temperatures between 93K and 333K. The current was measured between -1V and 1V in order to avoid the contribution of ferroelectric switching (the dynamic coercive voltage of BFO is around 5V). To verify that the measured currents are not due to relaxation/transient processes a preliminary investigation of current versus time I(t) has been performed. The results showed a stable current as a function of the time which confirms that the response is the real leakage current and excludes any contributions of trapped carriers or relaxation dipoles under the DC voltages. Fig.3 shows that the obtained current characteristics for temperatures higher than 253K present a small asymmetry (the current for positive and negative voltages are close but not identical) while for temperatures below 253K the I(V) characteristics are perfectly symmetric. In addition, above 253K a significant increase of the current with increasing temperature is observed. These results hint toward mixed mechanism of conduction in our BFO film depending on temperature. Interface limited transport mechanisms are known to be accompanied with asymmetric I(V) curves while symmetric I(V) points to a bulk limited transport mechanism. The asymmetry of I(V) characteristics for high temperature supports therefore the Schottky emission and for low temperatures (below 253K) the bulk mechanism seems to be the dominant mechanism. In reality both bulk and interface limited transport can coexist in the film and only a dominant transport mechanism in BFO film can be inferred from experimental investigations.

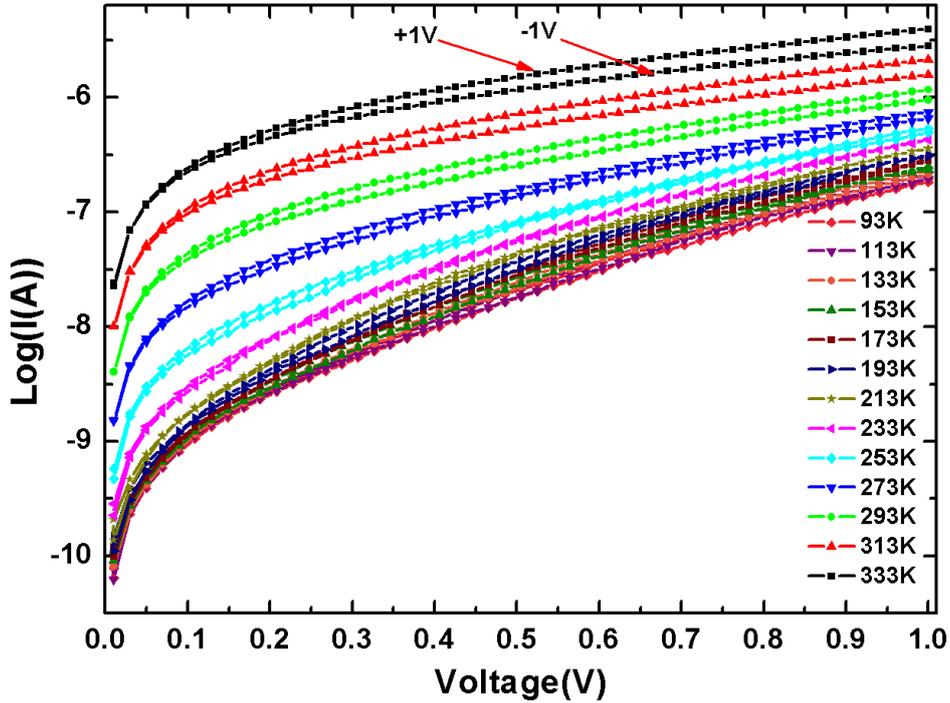

*Figure 3: log|I| versus voltage at different temperatures of the ITO/BFO/SRO capacitor.*

To understand the mechanism of conduction in ITO/BFO/SRO/STO heterostructure the models frequently used to describe carrier transport in ferroelectric thin films were considered. The I(V) characteristics were fitted to Schottky thermionic emission model and Poole-Frenkel conduction. The Schottky current density can be expressed as [14,19,20]:

$$J = A^*T^2 \exp\left[\frac{-q(\Phi_B - \sqrt{qE/4\pi\varepsilon_{op}\varepsilon_0})}{kT}\right] \qquad (1)$$

A* is the effective Richardson constant $A^* = \frac{120m^*}{m_0}$, $m_0$ the electron free mass, m* the electron effective mass, $\Phi_B$ the potential barrier at 0V, $\varepsilon_{op}$ the dynamic dielectric constant of the BFO, T the temperature and k the Boltzmann constant. The equation (1) can be written as:

$$\ln\left(\frac{J}{T^2}\right) = \ln(A^*) - \frac{q}{kT}(\Phi_{app}) \tag{2}$$

Where $\Phi_{app}$ is the apparent potential barrier

$$\Phi_{app} = \left(\Phi_B - \sqrt{qE/4\pi\varepsilon_{op}\varepsilon_0}\right) \tag{3}$$

A totally depleted film has been considered with the above relations and unrealistic $\varepsilon_{op}$ are obtained (see supporting information). Partially depleted film is thus discussed below. Considering a partially depleted film (field E is not uniform) the applied field can be written as [21]:

$$E = \sqrt{2qN_{eff}(V + V_{bi})/\varepsilon_{st}\varepsilon_0} \tag{4}$$

where $N_{eff}$ is the effective charge density at the depletion region, $\varepsilon_{st}$ is the static dielectric constant and $V_{bi}$ is an apparent built-in potential near the electrode/film interface. If $V_{bi} \ll V$, the $\Phi_{app}$ can be rewritten:

$$\Phi_{app} = \Phi_B - \sqrt{\frac{q}{4\pi\varepsilon_{op}\varepsilon_0}}\left(\frac{2qN_{eff}V}{\varepsilon_{st}\varepsilon_0}\right)^{1/4} \tag{5}$$

then the equation (2) can be expressed by:

$$\ln\left(\frac{J}{T^2}\right) = (\ln(A^*) - \frac{q}{kT}\Phi_B) + \left[\frac{q}{kT}\sqrt{\frac{q}{4\pi\varepsilon_{op}\varepsilon_0}}\left(\frac{2qN_{eff}}{\varepsilon_{st}\varepsilon_0}\right)^{1/4}\right]V^{1/4} \tag{6}$$

Figure 4 displays the Schottky representations (a) $\ln J/T^2$ versus $V^{1/4}$ and (b) $\ln J/T^2$ versus $1/T$. In agreement with a Schottky like mechanism, Fig. 4(a) shows a linear behaviour for voltages above 0.1V and temperatures above 273K. Note that for lower temperatures, Ln(J/T2) versus $V^{1/4}$ is not linear in this voltage range and the slope of the curves (for $V^{1/4} > 0.8\ V^{1/4}$) undergoes a drastic change compared to the high temperatures slopes (as shown by the linear fits plotted in Fig/4(a) for $V^{1/4} > 0.8\ V^{1/4}$). We also note that above 273K the I(V) characteristics (Fig.3) are asymmetric in full coherency with an interface limited transport of charges.

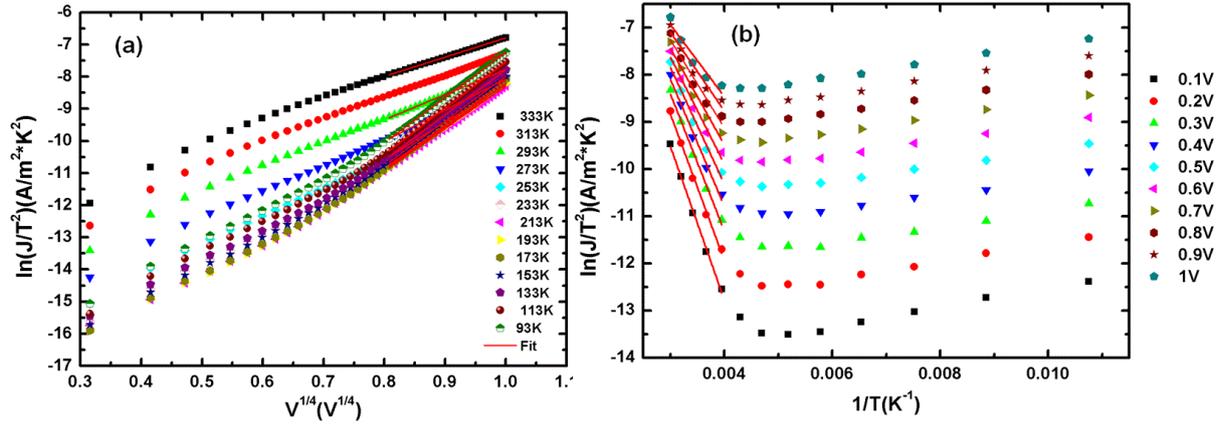

Figure 4: (a) ln(J/T$^2$) versus V$^{1/4}$ and (b) ln(J/T$^2$) versus 1/T Schottky representations. The solid lines in (a) and (b) show linear fits for V>0.4V and T>253K, respectively.

Fig. 4(b) shows two different regimes as a function of temperature. Linear fits for T>253K allow extracting $\Phi_{app}$ from the slope. From linear dependence of $\Phi_{app}$ versus V$^{1/4}$ we calculated the value of the potential barrier $\Phi_B$ (intercept for $V = 0$) and $N_{eff}$ (from the slope) considering $\varepsilon_{st}$=100 and $\varepsilon_{op}$=6.5 (typical values for BFO) [22,23]. These two parameters are found to be 0.63±0.01 eV and (1.76±0.02)*10$^{27}$ m$^{-3}$ respectively in agreement with values from the literature [24]. From the intercept at the origin of the fit of ln(J/T$^2$) versus 1/T representations we deduced a Richardson constant of 0.09±0.01 ≤A*≤1.25±0.04 A/m$^2$*K$^2$ which corresponds to 7.5*10$^{-4}$≤m*/m≤0.01. These values are much lower compared to values observed in perovskites (typically m*/m is 5–20). The small Richardson constant is explained by the low mobility of the charges within the bulk of the thin film and Schottky modified model (Schottky-Simmons) takes into account such behaviour with an exponential pre-factor depending on the electric field and mobility [21]. Therefore the obtained values of $\Phi_B$ and $N_{eff}$ strongly suggest an interface-limited transport mechanism described by a modified Schottky model. We ruled out tunneling at interface barriers since I(V) curves show a dependence with temperature in disagreement with a tunneling process. Next we present the contribution of bulk-limited mechanism by considering the Pool-Frenkel mechanism (PF). This mechanism corresponds to hopping of charges (electrons or holes) from defect trap centres to the conduction or valence band. In this case, the ionization of the trapped charges can be activated thermally or by an electric field. The current density of PF mechanism can be expressed by [14,25,26]:

$$J = q\mu N_c E \exp\left(\frac{-q\left(\Phi_T - \sqrt{\frac{qE}{\pi \varepsilon_{op}\varepsilon_0}}\right)}{kT}\right) \quad (8)$$

Where q is the elementary charge, μ the carrier mobility, $N_c$ the state density in conduction band, $\Phi_T$ the trap energy, $\varepsilon_{op}$ the dynamic dielectric constant. The equation (8) can be written as:

$$\ln\left(\frac{J}{E}\right) = \left(q\mu N_c - \frac{q}{kT}\Phi_T\right) + \frac{q}{kT}\sqrt{q/\pi\varepsilon_{op}\varepsilon_0}\sqrt{E} \qquad (9)$$

Figure 5 displays (a) ln(J/E) *versus* $E^{1/2}$ and (b) the dielectric constant $\varepsilon_{op}$ calculated from the linear fit of ln(J/E) versus $E^{1/2}$.

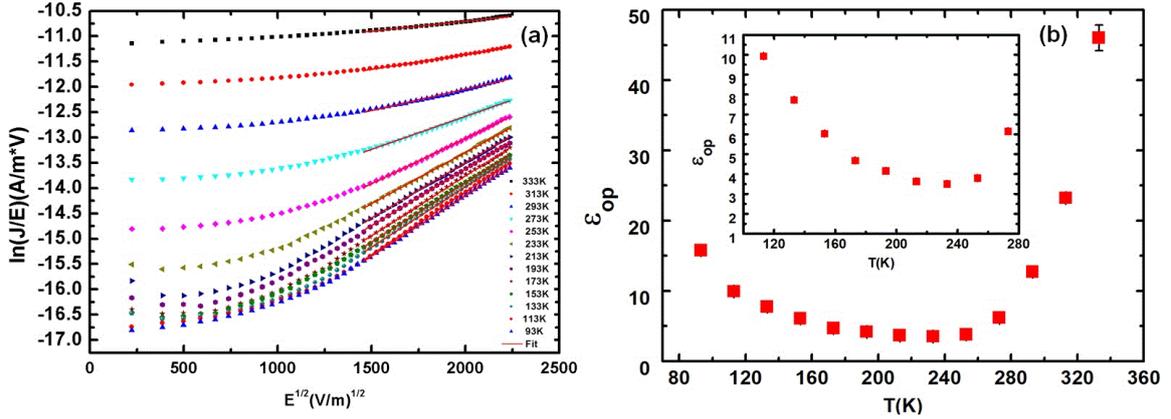

*Figure 5:* (a) $ln(J/T^2)$ versus $E^{1/2}$ and (b) the temperature evolution of the calculated dielectric constant $\varepsilon_{op}$. The solid lines in (a) show linear fits for $E^{1/2} \geq 1500(V/m)^{1/2}$.

We note that ln(J/E) *versus* $E^{1/2}$ (Fig.5(a)) shows a linear behaviour for high field ($E^{1/2} \geq 1500(V/m)^{1/2}$). The physical parameters such as $\varepsilon_{op}$ and $\Phi_T$ can thus be deduced from the slope and the intercept at the origin of the linear portion of ln(J/E) *versus* $E^{1/2}$. The extracted dielectric constant is plotted as a function of temperature in figure 5(b). For temperatures between 100 and 280 K (insert of Fig.5(b)) the $\varepsilon_{op}$ varies between 9.93±0.09 and 4.16±0.03. These values are close to usual reported values (6.5) for BFO [22,23]. For high temperatures however, $\varepsilon_{op}$ increases with temperature and reaches 46.00±1.80 for T=333K. The obtained values support a PF bulk contribution for the mechanism of conduction in the BFO film for the low temperature range. A change between two transport mechanisms is therefore inferred: a high temperature (T>253K) modified Schottky mechanism to a low temperature (T<253K) PF conduction mechanism.

The photovoltaic (PV) effect in (111) BFO film has been investigated by collecting J(V) curves under laser illumination (λ=514nm). The voltage was applied to the SRO bottom electrode and swept from +1V to -1V for room temperature and from +5V to -5V for temperature dependent measurements. Figure 6(a) displays room temperature (300K) J(V) characteristic recorded at different powers of laser illumination between 1 and 12mW. For all power laser illumination, a clear PV effect is observed with a negative open circuit voltage ($V_{oc}$) and positive short circuit current ($J_{sc}$) strongly dependent on the laser power. The dark J(V) curve crosses the origin and does not show any PV response. The evolution as a function of laser power of $V_{oc}$ and $J_{sc}$ is plotted in Fig. 6(b). The absolute value of $V_{oc}$ increases with the laser power to reach a value of 0.75V for 12mW laser power. $V_{oc}$ tends to saturate at high laser power and this is explained either by a local overheating or a favoured recombination of electron hole (above a certain threshold of photo-induced electron hole pairs). $J_{sc}$ increases linearly with increasing laser power from a value of 0.25mA/cm² for 1mW up to 3mA/cm² for 12mW laser power. The linear dependence of $J_{sc}$ implies that the number of photo-excited carriers in BFO film is determined by the laser power intensity [27].

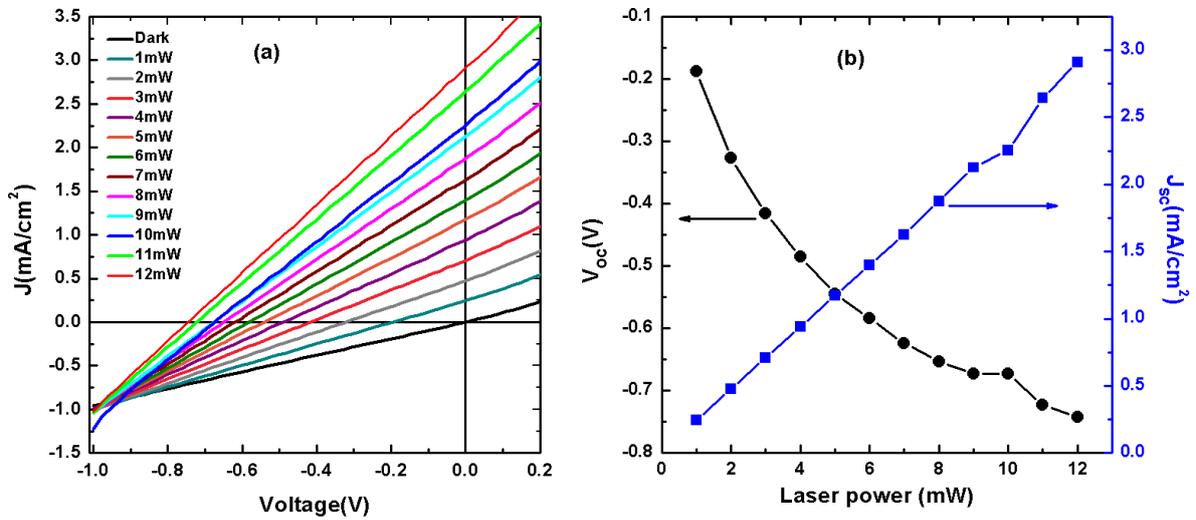

*Figure 6:(a) room temperature J(V) curves under different laser power illumination (at 514nm wavelength) showing PV effect (b) evolution of the $V_{oc}$ and $J_{sc}$ with the laser power.*

In ferroelectric thin films the measured PV effect can be related to different mechanisms such as the presence of an inhomogeneous potential at the electrode/film interfaces, the depolarizing field, electro-migration of vacancies or the non-centrosymmetric bulk PV effect [5,28,29]. All these mechanisms can contribute to the observed PV effect and it is challenging to identify the exact origin of the PV response. To evidence a possible influence of the ferroelectric polarization, J(V) curves were measured under light after application of voltage pulses (time duration of 90ms and 10mW laser power). The results are plotted in the figure 7(a). A maximum voltage pulse of +/- 6V superior to the coercive field of our BFO film is applied in order to investigate the effect of the observed polarization switching on the PV response (Fig.7a). The used protocol consists of applying first the pulse and then collect the J(V) curve by starting the sweeping voltage at same polarity than the pre-applied pulse, in order to avoid any back switching of induced polarized state prior to the J(V) measurement. For instance when pre-pulse of -2V is applied, the sweeping voltage for J(V) measurement starts on the negative side. A clear switchable PV effect is observed when we varied the pulses between -6V to 6V by step of 1V. The evolution of $V_{oc}$ and $J_{sc}$ with voltage pulses is shown in Fig.7(b). As mentioned before for the PV effect in the as-grown polarization state (0V pulse) the $V_{oc}$ is negative and when we applied a negative pulses up to -6V the absolute value of $V_{oc}$ shows a small increase and saturates around -0.55V. $J_{sc}$ is positive for negative pulses and saturates around 3mA/cm$^2$. For applied positive pulses the $V_{oc}$ and $J_{sc}$ remain almost constant until 2V then change abruptly to values close to 0 around +3.3V pulses. $V_{oc}$ ($J_{sc}$) is completely switched for V>+4V pulses and takes positive (negative) value. Thus, we suggest that the obtained PV effect under voltage pulses is dominated by the bulk ferroelectric polarization switching in BFO film. The anomaly observed at about -0.2V on the red curve of Fig. 7(a) and corresponding to the +6V pulse is explained by the back switching to the preferential orientation of the ferroelectric polarization (imprint). The positive imprint observed in the P-V loop when the positive voltage is applied on the SRO bottom electrode suggests a downward self ferroelectric polarization in agreement with the obtained PV effect for 0V pulse (negative $V_{oc}$ and positive $J_{sc}$) as shown in the insert of Fig.7 (b). Note that the value of 4V needed to fully switch the PV response is close to the coercive voltage of 5.3V observed in the dynamic P-V loop (figure 2).

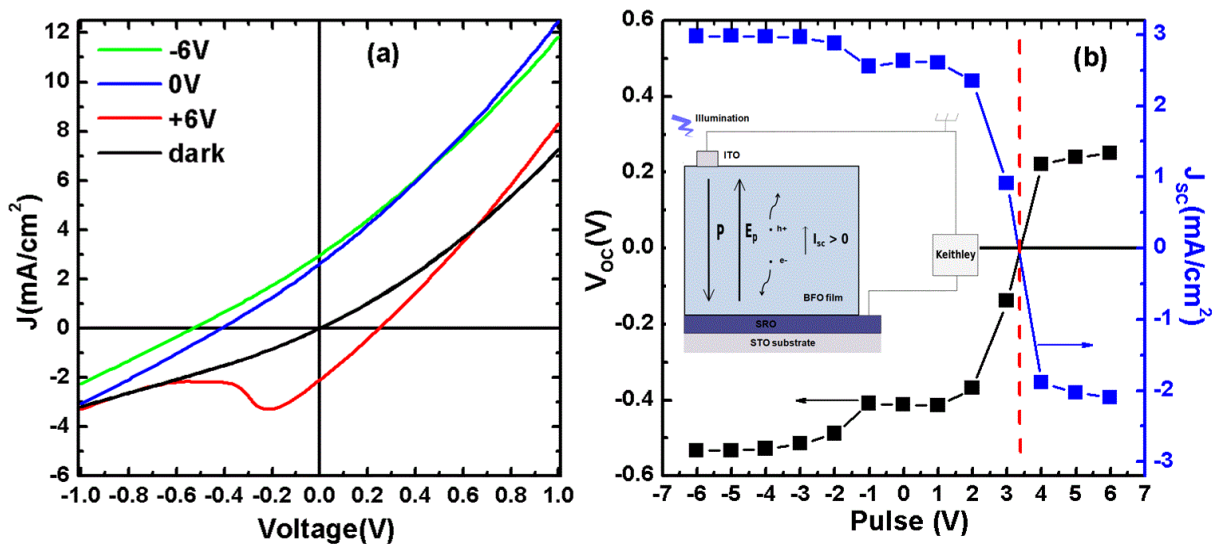

*Figure7: (a) room temperature J(V) curve under 10mW laser power (514nm) after different applied positive and negative pulses (b) evolution of $V_{oc}$ and $J_{sc}$ with applied voltage pulses. Insert in (b) shows the down orientation of the as grown ferroelectric polarization and the resulting positive $J_{sc}$.*

Recently Yousfi *et al*. investigated the ferroelectric properties and PV response of Pt/(100)BFO/SRO/LaAlO$_3$ heterostructure [15]. In this later investigation the imprint is negative with an upward self-polarization leading to a positive $V_{oc}$ and negative $J_{sc}$ in the as grown film. Interestingly, the results discussed here for the (111) BFO film grown in the similar conditions show that the PV response and conduction mechanism can be tuned by the orientation of the substrate and/or the type of top electrodes. To better understand the switchable PV effect in (111) BFO film we also explored the temperature dependence of the J(V) curves. Figure 8(a) shows J(V) characteristic curves recorded under 10mW laser power at different temperatures. The sweeping voltage is kept identical for all the measurements (from -5V to +5V). A strong dependence of PV effect as a function of temperature is observed in particular for the open circuit voltage. The temperature evolution of $V_{oc}$ and $J_{sc}$ are plotted in Fig.8 (b).

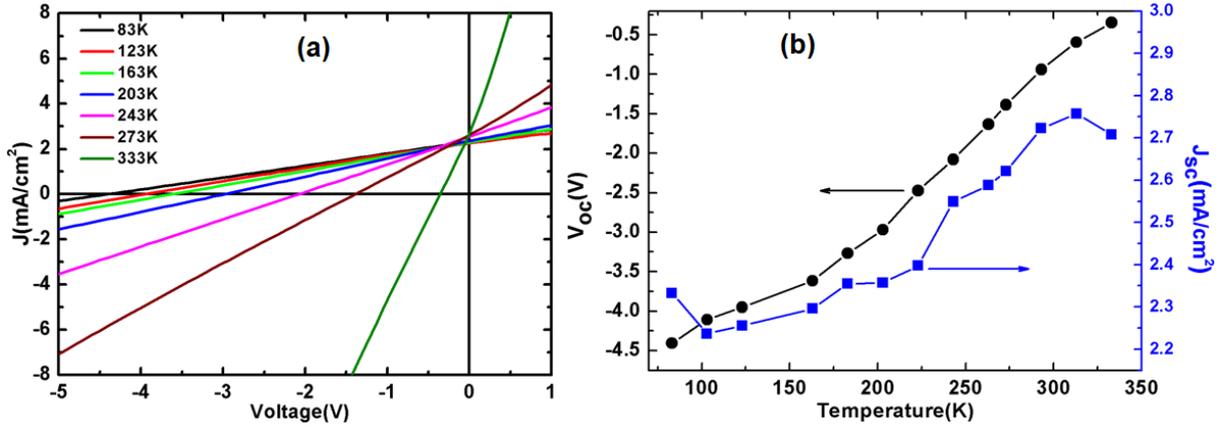

*Figure 8: (a) J(V) curve under 10mW laser power at different temperature (b) temperature dependence of $V_{oc}$ and $J_{sc}$.*

On cooling down the sample the absolute value of $V_{oc}$ increases and a large $V_{oc}$ of -4.5V is obtained at 83K. This behaviour allows us to rule out electromigration of oxygen vacancies as the main parameter influencing the PV response. Indeed a decrease of $V_{oc}$ is expected in this case on cooling as demonstrated by Ge *et al.* [30]. The large $V_{oc}$ confirms the dominant contribution of bulk like origin of the PV response while the small asymmetry between $V_{oc}$ values for up and down polarization state evidences a minor contribution of the Schottky barrier at the interfaces. Schottky barriers are indeed accompanied with $V_{oc}$ limited by the bandgap (2.7eV) [14]. Concerning the $J_{sc}$ dependence with temperature we believe that the small increase of the bandgap on cooling decreases the amount of photo-induced electron-hole pair and the subsequent decrease of $J_{sc}$. The impact of the change of transport mechanism evidenced above may also play a role and must not be ruled out.

**IV. Conclusion**

(111) oriented BiFe$_{0.95}$Mn$_{0.05}$O$_3$ thin film has been grown by pulsed laser deposition on (111) STO substrate buffered with 50 nm SRO. High-resolution x-ray diffraction characterization shows an epitaxial growth of the film without any second phase and a rhombohedral BFO bulk like structure. Using ITO as top electrode a saturated P-V loop with a large value of the remnant polarization of Pr=104μC/cm$^2$ is obtained. At high temperature the Schottky interface dominates the conduction mechanism that evolves to a Poole-Frenkel mechanism at low temperature. A switchable PV effect is detected at room temperature with a clear dominant ferroelectric origin over defects migration and interface effects. At low temperatures, above band-gap V$_{oc}$ value of -4.5V is observed strongly supporting the bulk non-centrosymmetric PV effect. Such results indicate the possibility to employ vertical geometry of measurements to engineer bulk-like PV effect with large V$_{oc}$ in multiferroic systems.

**Acknowledgement**

This work has been funded by the region Hauts-de-France (Projects ZOOM and NOTEFEV) and European Projects "NOTEDEV" FP7-People-ITN and H2020-MSCA-RISE "ENGIMA n° 778072".